\documentclass[a4paper,20pt]{article}
    \usepackage[T1]{fontenc}
    \usepackage{graphicx}
    \usepackage{epsfig}
    \usepackage{amsmath}
    \usepackage{amsfonts}
    \renewcommand{\abstract}{}
\begin{document}
\makeatletter
\renewcommand{\@oddhead}{\textit{YSC'14 Proceedings of Contributed Papers} \hfil \textit{S.Yu. Zubrin, A.V. Antyufeyev, V.M. Shulga}}
\renewcommand{\@evenfoot}{\hfil \thepage \hfil}
\renewcommand{\@oddfoot}{\hfil \thepage \hfil}
\fontsize{11}{11} \selectfont

\title{Methanol Masers Observations in the 3-mm Bandwidth at the Radio Telescope
RT-22 CrAO}
\author{\textsl{S.Yu. Zubrin$^{1}$, A.V. Antyufeyev, V.V. Myshenko, V.M. Shulga}}
\date{}
\maketitle
\begin{center} {\small Institute of Radio Astronomy, Kharkiv \\
$^{1}$zubrin@rian.kharkov.ua}
\end{center}

\begin{abstract}
We report the beginning of the astronomical masers investigations
in the 3-mm bandwidth at the radio telescope RT-22 (CrAO,
Ukraine). For this purpose the special complex for maser lines
investigation in 85...115 GHz frequency band is developed. It is
made on the base of the low noise cryogenic Shottky-diode receiver
and the high resolution Fourier-spectrometer. The cryogenic
receiver has the DSB noise temperature less than 100K. The
spectral channel separation of the Fourier-spectrometer is about
4kHz and the spectrometer bandwidth is 8 MHz. Results of maser
observations of 8$^{0}$-7$^{1} $A$^{ +} $ transition of methanol
(95.169 GHz) towards DR-21(OH), DR-21W and NGC7538 are in good
agreement with early obtained results by other authors. On the
basis of the analysis of the location of masers in the NGC7538
direction we can assume that the origin of all known class I
methanol masers in this region is connected with existing
molecular outflows from young stars.
\end{abstract}

\section*{Introduction}
\indent \indent Astronomical masers can be named a "point-like
probe" of the physical conditions and dynamical processes in places
of star burn. A strong interest to the investigation of thermal and
maser methanol radiation is caused by the rich methanol spectrum.
Multifrequency investigations of methanol masers are of particular
interest. Basically such investigations concern just two transitions
of methanol molecule 5$^{1}$-6$^{0}$ E (6.7 GHz) and
2$^{0}$-3$^{-1}$ E (12 GHz). Much more interesting are the methanol
radiation investigations in the 85\ldots115 GHz frequency band due
to 7 transitions revealing maser effect and 4 transitions revealing
thermal radiation that can be studied  in this frequency band
simultaneously. The only simultaneous investigations of all these
transitions of methanol were held towards 23 well known high-mass
star-forming regions \cite{min02} and give a global overview of
methanol masers in these directions. This permits us to make
conclusions about the relationship of different classes of methanol
masers, about the nature of maser sources and about the physical
conditions in these star-forming regions.

We report the beginning of methanol masers investigations at the
radio telescope RT-22 (CrAO) in 85\ldots115 GHz frequency band.

\section*{Complex for maser lines investigations}

\begin{figure}[h]
\centerline{%
\begin{tabular}{c@{\hspace{0.5in}}c}
\includegraphics[width=2.4in]{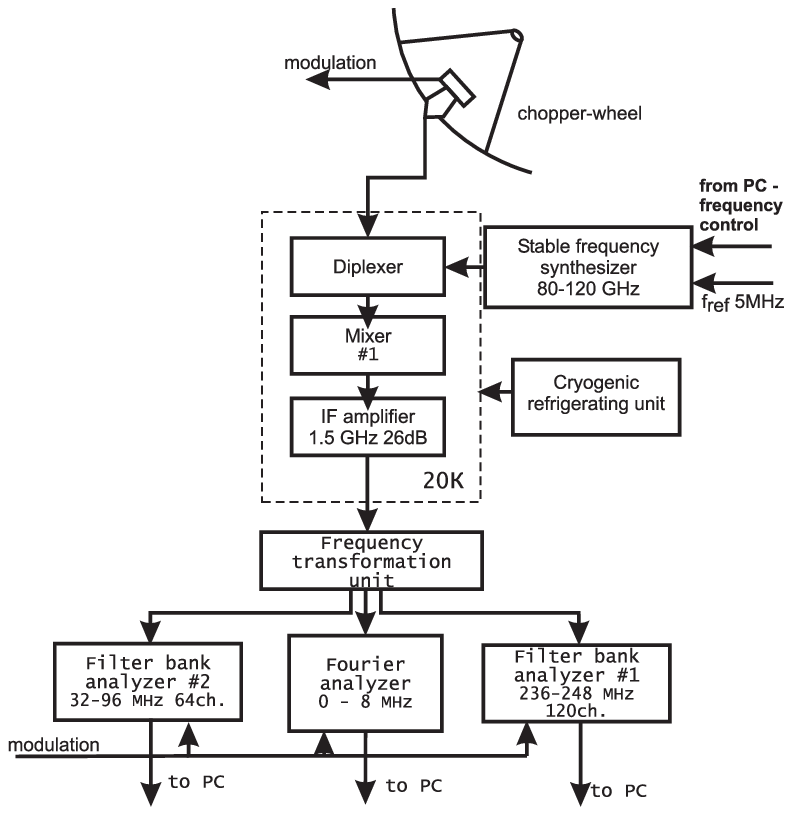} &
\includegraphics[width=2.4in]{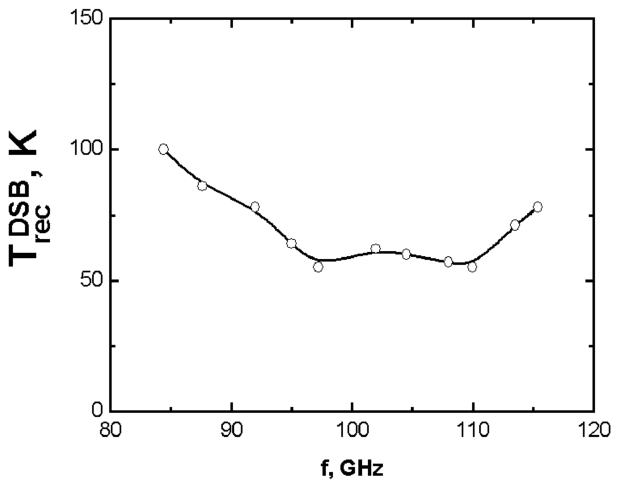} \\
a.~ & b.~
\end{tabular}}
\begin{center}
\caption{\small The block diagram of the complex for maser lines
 investigations (a) and the noise temperature of the cryogenic receiver in
 the frequency band 85\ldots115 GHz in the DSB regime (b).}
\end{center}
\end{figure}

\indent \indent Special complex for study of maser lines of space
sources at the radio-telescope RT-22 was made on the base of high
sensitive cryogen superheterodyne 3 mm receiver and three spectrum
analyzers. A Fourier spectrum analyzer with high frequency
resolution was used in this complex. The block diagram of complex is
shown on Fig.1a.

Cryogen receiver presents the cooled part of the complex. The mixer
diode chip placed in half-height waveguide. To combine RF and LO
signal in receiver diplexer unit is used, whose additional function
is 'cleaning' the LO spectrum. The LO is PL L synthesizer with BWT
OB-71. Cooled IF amplifier is designed at IRA NASU, used 2 PHEMT
transistors 'Agilent' and provides gain more than 20dB and noise
temperature less then 2K under 20K ambient temperature \cite{kor04}.
Diplexer, mixer and IF amplifier were placed in the cryoblock and
cooled up to 20K with a help of closed circle criogenic cooler.
Noise temperature of receiver was measured in laboratory and its
value (T$_{N}^{DSB} $) was less then 100K at any frequency in the
85\ldots115 GHz band (see Fig. 1b).

Uncooled part of the complex is used for amplification of received
signal on intermediate frequency and for coupling with 2 types of
spectrum analyzer. One of them is filter bank spectrum analyzers
with low frequency resolution. It increase the frequency band of
the complex up to 12 MHz. Fourier analyzer can operate in 2
regimes and provides our complex with high frequency resolution
(3.9 kHz).

\section*{Methanol masers observations.}
\indent \indent Observations were carried out on November, 2004 at
the 22m radio telescope RT-22 (CrAO, Simeiz). The effective area of
the radio telescope was estimated with planets (Jupiter, Venus)
observations and at 95 GHz it was 65 m$^{2}$. The main beam width of
the radio telescope was estimated from the planet observations and
was $40''$. The root-mean-square pointing error of the radio
telescope was $1''$.

During the observations three maser sources were detected towards
NGC7538, DR21(OH), DR21W. The results of observations are presented
in Table 1 and they are in good agreement with earlier results
retrieved by other authors  \cite{pla90, kal94}.

\begin{table}[h]
{\footnotesize
\caption{\small Line parameters, determined from
CH$_{3}$OH 8$^{0}$-7$^{1} $A$^{ +} $ masers observations}
\label{tab1}
\begin{center}
\begin{tabular}{cccccc}
\hline\noalign{\smallskip}
 $Source$ & $\alpha _{2000}$ & $\delta_{2000}$ & $V_{lsr},\ km/sec$ &
 $\Delta V_{lsr},\ km/sec$ & $S,\ Jy$ \\
\noalign{\smallskip} \hline \noalign{\smallskip}
 NGC7538 & 23$^{h}$13$^{m}$46.5$^{s}$ & +61\r{} 27'\ 32.7'' & -57.44& 0.75 & 21 \\
 DR21(OH)& 20$^{h}$38$^{m}$59.2$^{s}$ & +42\r{} 22'\ 48.7'' & +0.16 & 0.73 & 136 \\
 DR21-W  & 20$^{h}$38$^{m}$54.6$^{s}$ & +42\r{} 19'\ 23.5'' & -2.69 & 1.18 & 52 \\
\noalign{\smallskip} \hline
\end{tabular}
\end{center}
}
\end{table}

DR21(OH) and DR21W.

We detect maser sources towards DR21: DR21(OH) and DR21W. The
observed spectra are shown in Fig. 2. This maser sources were
earlier detected on the BIMA by the authors of \cite{pla90} and
their line width were twice narrower than in our observations. This
fact in the case of DR21(OH) can be explained by the simultaneous
observations of two masers DR21(OH)-1 and DR21(OH)-2 \cite{pla90} ,
that can be resolved during interferometric observations and can not
be resolved on a single-dish telescope. The earlier observations of
this maser sources on a single-dish telescope \cite{kal94} reveal
the same maser line parameters as in our observations.

NGC7538.

NGC7538 is an HII region and a young-stars rich cluster. There are
11 IR sources (marked IRS1-11) towards. Moreover 2 bipolar outflows
from IRS1 and IRS11 can be seen in this region during CS and CO
observations \cite{kam89}. Several 44 GHz methanol masers were found
in this region \cite{kur04}. The map of these outflows is shown in
Fig. 3, also here are all 44 GHz and 95 GHz methanol masers towards
this direction. It can be seen, that 44 GHz and 95 GHz methanol
masers are placed between blue-shifted wing of the outflow from
IRS11 (-64\ldots-62 km/sec) and red-shifted wing of the outflow from
IRS1 (-45\ldots-40 km/sec). The both radial velocities of detected
95 GHz methanol maser (-57.44 km/sec) and of detected 44 GHz
methanol masers (-56.6\ldots-59.2 km/sec) agree with radial
velocities of methanol thermal lines at 133 and 157 GHz in this
direction (-57.1\ldots-58.7 km/sec) \cite{sly99}.

As long as both 95 and 44 GHz masers belongs to masers with
collisional pumping \cite{sob93} and their velocities sufficiently
good agree with velocities of the wings of the outflows presented
here we assume that these masers originate from interaction of these
outflows and molecular cloud.

\begin{figure}[p]
\centerline{%
\begin{tabular}{c@{\hspace{0.3in}}c}
\includegraphics[width=3.3in]{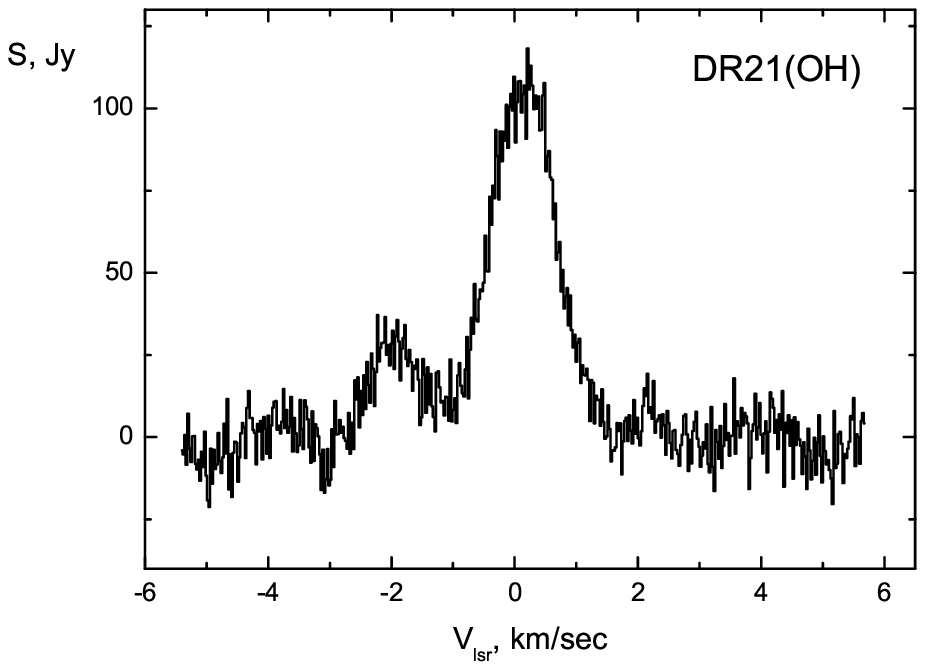} &
\includegraphics[width=3.3in]{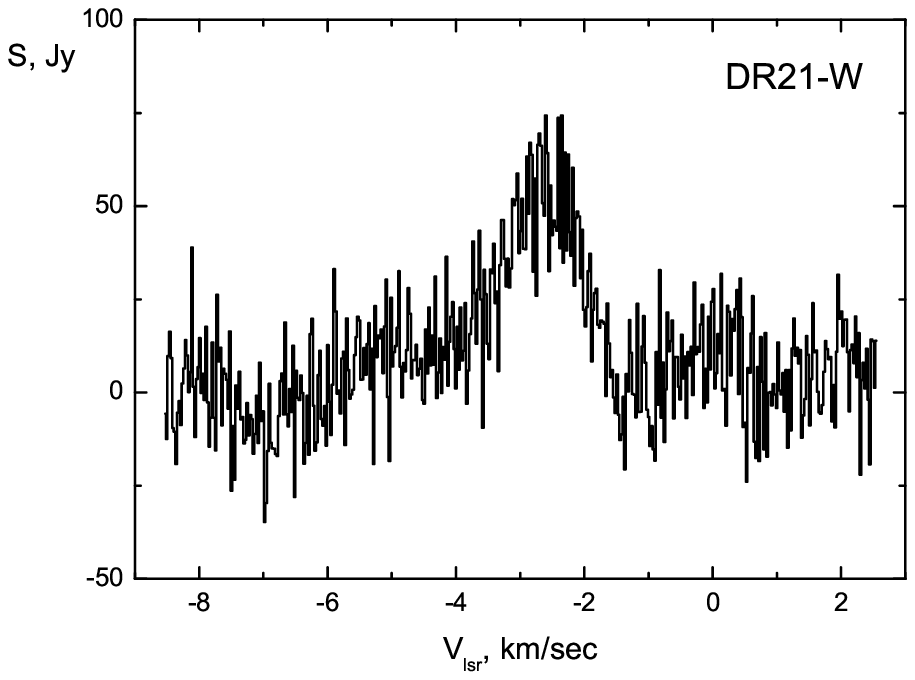} \\
\end{tabular}}
\caption{\small Spectra of maser sources towards DR21(OH) and
DR21W.}
\end{figure}
\begin{figure}[p]
\centerline{%
\begin{tabular}{c@{\hspace{0.5in}}c}
\includegraphics[width=3.3in]{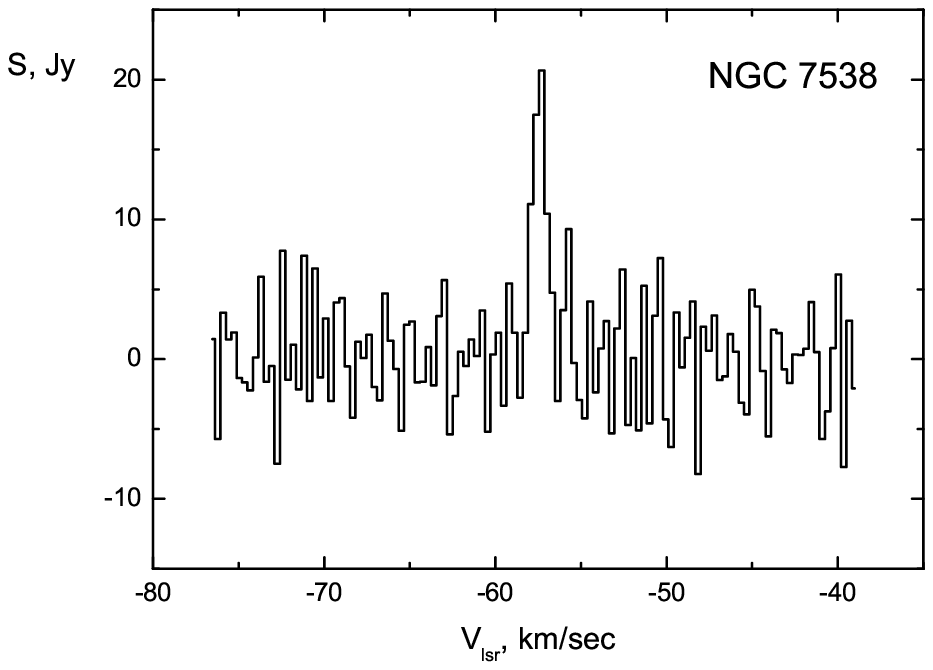} &
\includegraphics[width=2.2in]{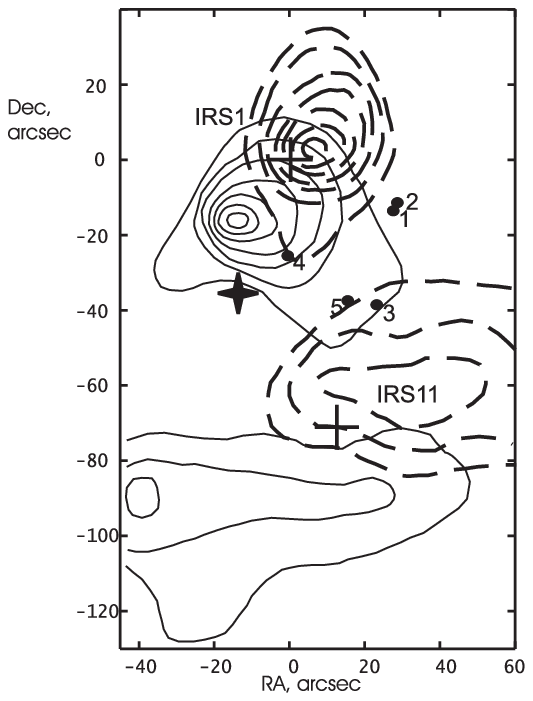} \\
a.~ & b.~
\end{tabular}}
\begin{center}
\caption{\small The spectrum of the maser source towards NGC7538 (a)
and the map of the molecular outflows from IRS1 and IRS11 towards
NGC7538 in CO-line \cite{kam89} with class I methanol masers
(7$^{0}$-6$^{1}$A$^{+}$ (44GHz) masers \cite{kur04} are marked with
black circles, with a star marked the detected maser
8$^{0}$-7$^{1}$A$^{+}$ (95GHz)) (b). IR sources IRS1 and IRS11 are
marked with crosses. Red- and blue-shifted wings of outflows are
marked with solid and dashed lines accordingly. Position (0,~0) of
the map corresponds to the position of IRS1 (23$^{h }$13$^{m
}$45.5$^{s }$,
 +61\r{}28'\ 09'', J2000)}
\end{center}
\end{figure}

\section*{Conclusions.}
\indent \indent With the low noise cryogenic Shottky-diode receiver
and high resolution Fourier-spectrometer the high sensitive complex
for maser lines investigations with frequency resolution 3.9 kHz in
85\ldots115 GHz frequency band at the radio telescope RT-22 (CrAO)
was developed and first observations were carried out.

Relying on the results of observations of methanol masers towards
NGC7538 we assume that all class I methanol masers in this
direction originate from the interaction of existing here bipolar
outflows from IR sources IRS1 and IRS11 with molecular cloud.

\end{document}